# Deciphering Spectral Fingerprints of Habitable Extrasolar Planets

L. Kaltenegger[1], F. Selsis[2], M. Fridlund[3], H. Lammer[4], Ch. Beichman[5], W. Danchi[6], C. Eiroa[7], T. Henning[8], T. Herbst[8], A. Léger[9], R. Liseau[10], J. Lunine[11], F. Paresce[12], A. Penny[13], A. Quirrenbach[14], H. Röttgering[15], J. Schneider[16], D. Stam[17], G. Tinetti[18], G. J. White[13,19]

[1]Harvard University /Harvard Smithsonian Center for Astrophysics, Cambridge, MA, USA
[2]University of Bordeaux 1, Bordeaux, France, [3]RSSD, ESA, ESTEC, Noordwijk, The Netherlands
[4]Space Research Institute, Austrian Academy of Sciences, Graz, Austria, [5]NASA ExoPlanet Science Institute, California Inst. Of Technology/JPL, USA, [6]Goddard Space Flight Center, Greenbelt, MD, USA, [7]Universidad Autonoma de Madrid, Madrid, Spain, [8]Max-Planck Institut für Astronomie, Heidelberg, Germany, [9]Universite Paris-Sud, Orsay, France, [10]Dept. of Radio and Space Science, Chalmers University of Technology, Onsala, Sweden, [11]Lunar and Planetary Laboratory, University of Arizona, USA, [12]INAF,Via del Parco Mellini 84, Rome, Italy, [13]Space Science & Technology Dep., CCLRC Rutherford Appleton Laboratory, Oxfordshire, UK, [14]Landessternwarte, Heidelberg, Germany, [15]Leiden Observatory, Leiden, The Netherlands, [16]Observatoire de Paris-Meudon, LUTH, Meudon, France, [17]SRON, Netherlands Institute for Space Research, Utrecht, The Netherlands, [18]Department of Physics and Astronomy, University College London, London, UK, [19]The Open University, Milton Keynes, UK

Corresponding Author: L. Kaltenegger, CfA, Cambridge, MA, USA
E-mail: lkaltene@cfa.harvard.edu

**Abstract:**
In this paper we discuss how we can read a planet's spectrum to assess its habitability and search for the signatures of a biosphere. After a decade rich in giant exoplanet detections, observation techniques have now reached the ability to find planets of less than 10 $M_{Earth}$ (so called Super-Earths) that may potentially be habitable. How can we characterize those planets and assess if they are habitable? The new field of extrasolar planet search has shown an extraordinary ability to combine research by astrophysics, chemistry, biology and geophysics into a new and exciting interdisciplinary approach to understand our place in the universe. The results of a first generation mission will most likely result in an amazing scope of diverse planets that will set planet formation, evolution as well as our planet in an overall context.

Keywords: Extrasolar Planets, Biomarkers, Planetary atmospheres, Spectroscopy

## Introduction

Sagan et al. (1993) analyzed a spectrum of the Earth taken by the Galileo probe, searching for signatures of life and concluded that the large amount of $O_2$ and the simultaneous presence of $CH_4$ traces are strongly suggestive of biology. To characterize a planet's atmosphere and its potential habitability, we look for absorption features in the emergent and transmission spectrum of the planet. The spectrum of the planet can contain signatures of atmospheric species, what creates its spectral fingerprint. On Earth, some atmospheric species exhibiting noticeable spectral features in the planet's spectrum result directly or indirectly from biological activity: the main ones are $O_2$, $O_3$, $CH_4$, and $N_2O$. $CO_2$ and $H_2O$ are in addition important as greenhouse gases in a planet's atmosphere and potential sources for high $O_2$ concentration from photosynthesis.

The detection of Earth-like planet is approaching rapidly thanks to radial velocity surveys (HARPS), transit searches (Corot, Kepler) and space observatories dedicated to their characterization are already in development phase (James Webb Space Telescope), as well as large ground based telescopes (ELT, TNT, GMT), and dedicated space-based missions (Darwin, Terrestrial Planet Finder, New World Observer). In the next year, space missions like CoRoT (CNES, (Rounan et al. 1998)) and Kepler (NASA, (Bourucki et al. 1997)) will give us statistics on the number, size, period and orbital distance of planets, extending to terrestrial planets on the lower mass range end as a first step, while future space missions are designed to characterize their atmospheres. After a decade rich in giant exoplanet detections, indirect ground based observation techniques have now reached the ability to find planets of less than 10 $M_{Earth}$ (so called Super-Earths) around small stars that may potentially be habitable see e.g (Mayor et al. 2009)(Valencia et al. 2007). These planets can be characterized with future space missions.

The current status of exoplanet characterization shows a surprisingly diverse set of giant planets. For a subset of these, some properties have been measured or inferred using observations of the host star, a background star, or the combination of the star and planet photons (radial velocity (RV), micro-lensing, transits, and astrometry). These observations have yielded measurements of planetary mass, orbital elements and (for transits) the planetary radius and during the last few years, physical and chemical characteristics of the upper atmosphere of some of the transiting planets. Specifically, observations of transits, combined with RV information, have provided estimates of the mass and radius of the planet (see e.g., (Torres et al. 2008)), planetary brightness temperature (Charbonneau et al. 2005)(Deming et al. 2005), planetary day-night temperature difference (Harrington et al. 2006)(Knutson et al. 2007), and even absorption features of giant planetary upper-atmospheric constituents: sodium (Charbonneau et al. 2002), hydrogen (Vidal-Madjar et al. 2004), water (Swain et al. 2008)(Tinetti et al. 2007), methane (Swain et al. 2008), carbon monoxide and dioxide (Swain et al. 2008). The first imaged exoplanet candidates around young stars show the improvement in direct detection techniques that are designed to resolve the planet and collect its photons. This can currently be achieved for widely separated young objects and has already detected exoplanets candidates (see e.g. Kalas et al. 2008, Lagrange et al. 2009, Marois et al. 2008). Future space missions have the explicit purpose of detecting other Earth-like worlds, analyzing their characteristics, determining the composition of their atmospheres, investigating their capability to sustain life as we know it, and searching for signs of life. They also have the capacity to investigate the physical properties and composition of a broader diversity of planets, to understand the formation of planets and interpret potential biosignatures. Figure 2 shows the detectable features in the planet's reflection, emission and transmission spectrum using the Earth itself as a proxy.

In this paper we discuss how we can read a planet's spectral fingerprint and characterize if it is potentially habitable. In section 1 we discuss the first steps to detect a habitable planet and set biomarker detection in context. Section 2 focuses on low resolution biomarkers in the spectrum of an Earth-like planet, in section 3 we discuss spectral evolution of a habitable planet, cryptic worlds, abiotic sources of biomarkers, and Earth's spectra around different host stars, and section 4 summarizes the chapter.

## 1 Characterize a habitable planet

A planet is a very faint, small object close to a very bright and large object, its parent star. In the visible part of the spectrum we observe the starlight, reflected off the planet, in the IR we detect the planets own emitted flux. The Earth-Sun intensity ratio is about $10^{-7}$ in the thermal infrared (~10 μm), and about $10^{-10}$ in the visible (~0.5 μm) (see figure 1), but the contrast ratio of hot extrasolar Giant planets (EGP) to their parent stars flux as well as the contrast ratio of a planet to a smaller parent star is much more favorable, making Earthlike planets around small stars very interesting targets.

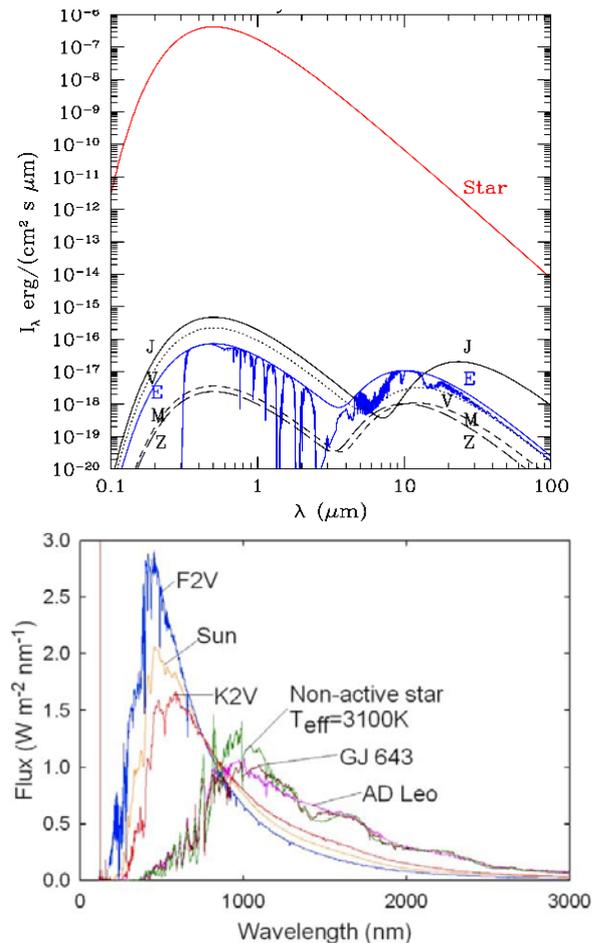

**Fig. 1:** SAO model of our Solar System (top) (assumed here to be Black Bodies with Earth spectrum shown). Spectra of different host stars (Segura et al 2005) (bottom panel).

The spectrum of the planet can contain signatures of atmospheric species, what creates its spectral fingerprint. The tradeoff between contrast ratio and design is not discussed here, but lead to several different configurations for space-based mission concept. Figure 2 shows observations and model fits to spectra of the Earth in 3 wavelength ranges (Kaltenegger et al 2007). The data shown in figure 2 (left) is the visible Earthshine spectrum (Woolf et al. 2002), (center) is the near-infrared Earthshine spectrum (Turnbull et al. 2006), and (right) is the thermal infrared spectrum of Earth as measured by a spectrometer enroute to Mars (Christensen & Pearl 1997). The data are shown in black the SAO model in red. In each case the constituent gas spectra in a clear atmosphere are shown in the bottom panel, for reference.

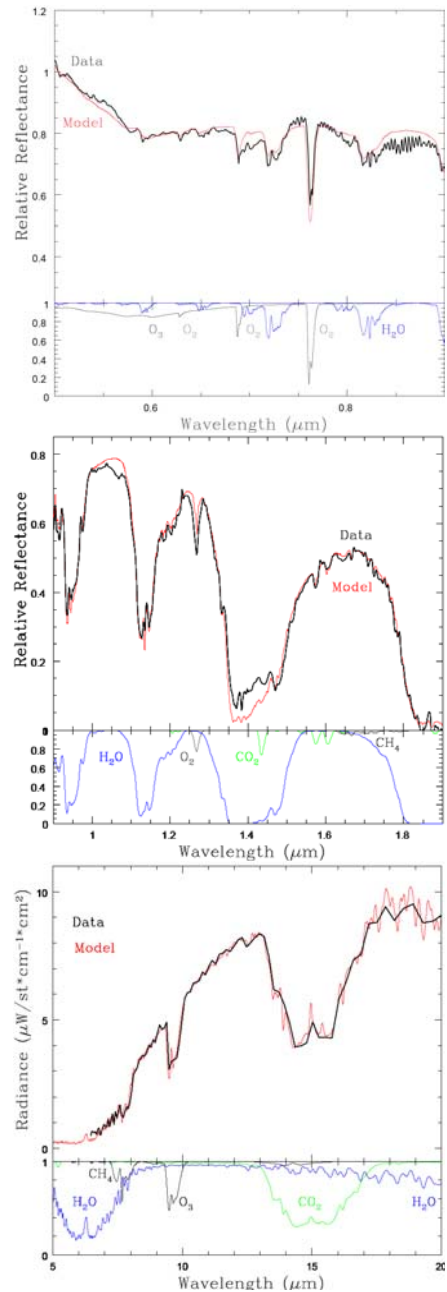

**Fig. 2:** Observed reflectivity spectrum in the visible (Woolf et al. 2002) (top), near-infrared (Turnbull et al. 2006) (middle) and emission spectrum in the infrared (Christensen & Pearl 1997) (bottom panel) of the integrated Earth, as determined from Earthshine and space respectively. The data is shown in black and the SAO model in red. The reflectivity scale is arbitrary.

The interferometric systems suggested operate in the mid-IR (6-20 μm) and observe the thermal emission emanating from the planet. The coronagraph and occulter concepts detect the reflected light of a planet and operate in the visible and near infrared (0.5-1μm). The viewing geometry results in different flux contribution of the overall detected signal from the bright and dark side, for the reflected light, and the planet's hot and cold regions for the emitted flux.

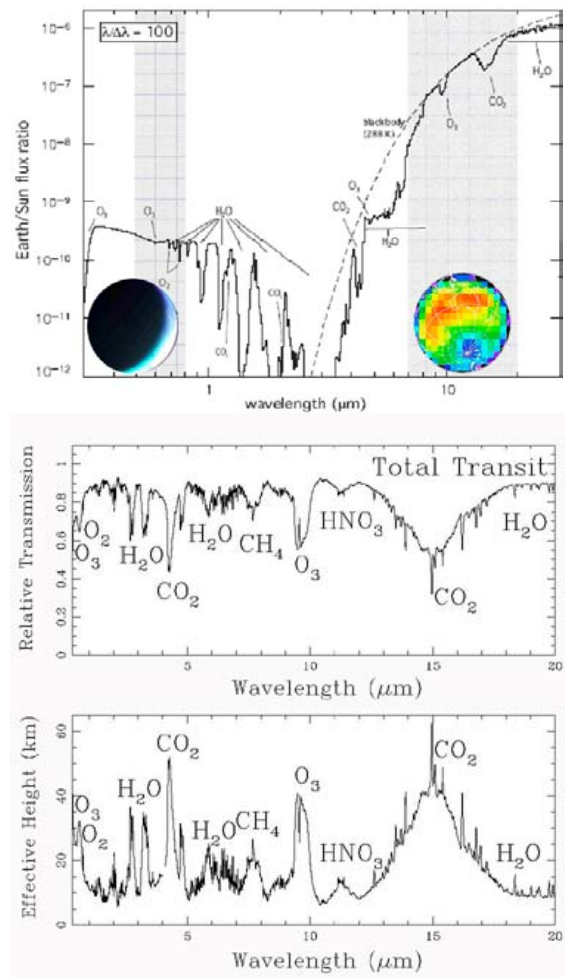

**Fig. 3:** Synthetic reflection and emission spectra (top) and transmission spectra (bottom panel) of the Earth from UV to IR shown. The intensity is given as a fraction of solar intensity as well as the relative height in the atmosphere. The atmospheric features are indicated.

Both spectral regions contain the signature of atmospheric gases that may indicate habitable conditions and, possibly, the presence of a biosphere: $CO_2$, $H_2O$, $O_3$, $CH_4$, and $N_2O$ in the thermal infrared, and $H_2O$, $O_3$, $O_2$, $CH_4$ and $CO_2$ in the visible to near-infrared. The presence or absence of these spectral features (detected individually or collectively) will indicate similarities or differences with the atmospheres of terrestrial planets, and its astrobiological potential (figure 3, see Palle et al. 2009 and Kaltenegger & Traub 2009 for details on Earth's transmission spectrum).

Our search for signs of life is based on the assumption that extraterrestrial life shares fundamental characteristics with life on Earth, in that it requires liquid water as a solvent and has a carbon-based chemistry (see e.g. Brack 1993, DesMarais et al. 2002). Life on the base of a different chemistry is not considered here because the vast possible life-forms produce signatures in their atmosphere that are so far unknown. Therefore we assume that extraterrestrial life is similar to life on Earth in its use of the same input and output gases, that it exists out of thermodynamic equilibrium (Lovelock 1975). Biomarkers is used here to mean detectable species, or set of species, whose presence at significant abundance strongly suggests a biological origin (e.g. couple $CH_4+O_2$, or $CH_4+O_3$, (Lovelock 1975)). Bio-indicators are indicative of biological processes but can also be produced abiotically. It is their quantities, and detection along with other atmospheric species, and in a certain context (for instance the properties of the star and the planet) that points toward a biological origin.

**1.1 Characterizing planetary environment**

It is relatively straightforward to remotely ascertain that Earth is a habitable planet, replete with oceans, a greenhouse atmosphere, global geochemical cycles, and life – if one has data with arbitrarily high signal-to-noise and spatial and spectral resolution. The interpretation of observations of other planets with limited signal-to-noise ratio and spectral resolution as well as absolutely no spatial resolution, as envisioned for the first generation instruments, will be far more challenging and implies that we need to gather information on the planet environment to understand what we will see.

The following step by step approach can be taken to set the planetary atmosphere in context. After detection, we will focus on main properties of the planetary system, its orbital elements as well as the presence of an atmosphere using the light curve of the planet or/and a crude estimate of the planetary nature using very low-resolution information (3 or 4 channels). Then a higher resolution spectrum will be used to identify the compounds of the planetary atmosphere, constrain the temperature and radius of the observed exoplanet. In that context, we can then test if we have an abiotic explanation of all compounds seen in the atmosphere of such a planet. If we do not, we can work with the exciting biotic hypothesis. $O_2$, $O_3$, $CH_4$ are good biomarker candidates that can be detected by a low-resolution (Resolution < 50) spectrograph. Note that if the presence of biogenic gases such as $O_2/O_3$ + $CH_4$ may imply the presence of a massive and active biosphere, their absence does not imply the absence of life. Life existed on Earth before the interplay between ocygenic photosynthesis and carbon cycling produced an oxygen-rich atmosphere.

## 1.1 Temperature and Radius of a Planet

Knowing the temperature and planetary radius is crucial for the general understanding of the physical and chemical processes occurring on the planet (tectonics, hydrogen loss to space). In theory, spectroscopy can provide some detailed information on the thermal profile of a planetary atmosphere. This however requires a spectral resolution and a sensitivity that are well beyond the performance of a first generation spacecraft. Here we concentrate on the initially available observations here.

One can calculate the stellar energy of the star $F_{star}$ that is received at the measured orbital distance. The surface temperature of the planet at this distance depends on its albedo and on the greenhouse warming by atmospheric compounds. However, with a low resolution spectrum of the thermal emission, the mean effective temperature and the radius of the planet can be obtained. The ability to associate a surface temperature to the spectrum relies on the existence and identification of spectral windows probing the surface or the same atmospheric levels. Such identification is not trivial. For an Earth-like planet there are some atmospheric windows that can be used in most of the cases, especially between 8 and 11 μm as seen in figure 3. This window would however become opaque at high $H_2O$ partial pressure (e.g. the inner part of the Habitable Zone (HZ) where a lot of water is vaporized) and at high $CO_2$ pressure (e.g. a very young Earth or the outer part of the HZ).

The accuracy of the radius and temperature determination will depend on the quality of the fit (and thus on the sensitivity and resolution of the spectrum), the precision of the Sun-star distance, the cloud coverage and also the distribution of brightness temperatures over the planetary surface. Assuming the effective temperature of our planet were radiated from the uppermost cloud deck at about 12km would introduce about 2% error on the derived Earth radius. For transiting planets, for which the radius is known, the measured IR flux can directly be converted into a brightness temperature that will provide information on the temperature of the atmospheric layers responsible for the emission. If the mass of non-transiting planets can be measured (by radial velocity and/or astrometric observations), an estimate of the radius can be made by assuming a bulk composition of the planet which can then be used to convert IR fluxes into temperatures. Important phase-related variations in the planet's flux are due to a high day/night temperature contrast and imply a low greenhouse effect and the absence of a stable liquid ocean. Therefore, habitable planets can be distinguished from airless or Mars-like planets by the amplitude of the observed variations of $T_b$ (see figure4).

The orbital flux variation in the IR can distinguish planets with and without an atmosphere in the detection phase (see also Selsis 2003, Gaidos et al. 2004). Strong variation of the thermal flux with the phase reveals a strong difference in temperature between the day and night hemisphere of the planet, a consequence of the absence of a dense atmosphere. In such a case, estimating the radius from the thermal emission is made difficult because most of the flux received comes from the small and hot substellar area.

The ability to retrieve the radius in such would depend on the assumption that can be made on the orbit geometry and the rotation rate of the planet. In most cases, degenerate solutions will exist. When the mean brightness temperature is stable along the orbit, the estimated radius is more reliable. The radius can be measured at different points of the orbit and thus for different values of $T_b$, which should allow to estimate the error made.

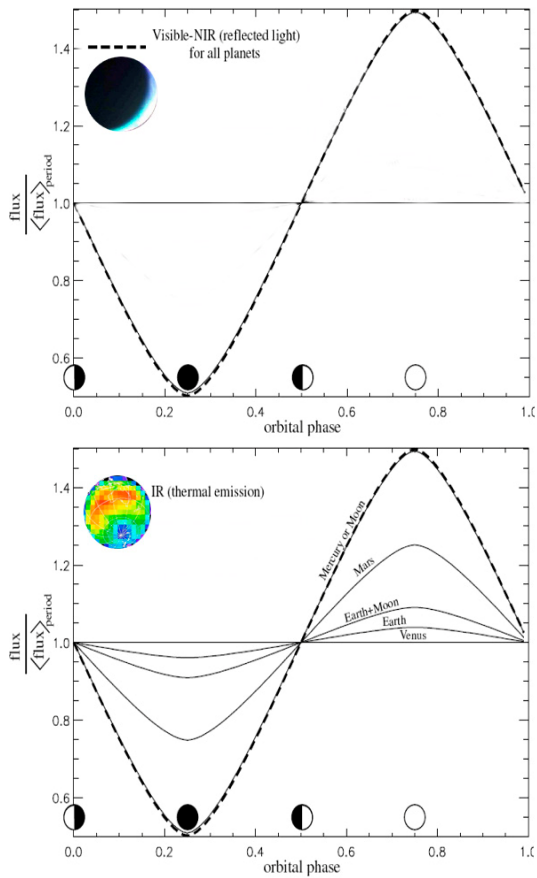

**Fig. 4: Orbital light curve for black body planets in a circular orbit with null obliquities, with and without an atmosphere in the visible (top) and thermal infrared (bottom) (after Selsis et al 2003).**

Note that also a Venus-like exoplanet would exhibit nearly no measurable phase-related variations of their thermal emission, due to the fast rotation of its atmosphere and its strong greenhouse effect and can only be distinguished through spectroscopy from habitable planets. The mean value of $T_b$ estimated over an orbit can be used to estimate the albedo of the planet, A, through the balance between the incoming stellar radiation and the outgoing IR emission.

The thermal light curve (i.e. the integrated infrared emission measured at different position on the orbit) exhibits smaller variations due to the phase (whether the observer sees mainly the day side or the night side) and to the season for a planet with an atmosphere than the corresponding visible lightcurve (see figure 4). In the visible ranges, the reflected flux allows us to measure the product $A \times R^2$, where R is the planetary radius (a small but reflecting planet appears as bright as a big but dark planet). The first generation of optical instruments will be very far from the angular resolution required to directly measure an exoplanet radius.

Presently, radius measurements can only be performed when the planet transits in front of its parent star, by an accurate photometric technique. If the secondary eclipse of the transiting planet can be observed (when the planet passes behind the star), then the thermal emission of the planet can be measured, allowing the retrieval of mean brightness temperature $T_b$ thanks to the knowledge of the radius from the primary transit. If a non-transiting target is observed in both visible and IR ranges, the albedo can be estimated in the visible once the radius is inferred from the IR spectrum, and compared with one derived from the thermal emission only.

## 2   Habitable Planets

The circumstellar Habitable Zone (HZ) is defined as the region around a star within which starlight is sufficiently intense to maintain liquid water at the surface of the planet, without initiating runaway greenhouse conditions vaporizing the whole water reservoir and, as a second effect, inducing the photodissociation of water vapor and the loss of hydrogen to space (see e.g. Kasting et al. 1993, 1997, Selsis 2000, Kaltenegger & Segura 2009 for a detailed discussion). The semi-major axis in the middle of the habitable zone $a$(HZ, AU), is derived by scaling the Earth-Sun system using $L_{star}/L_{sun} = (R_{star}/R_{sun})^2 (T_{star}/T_{sun})^4$, so

$a_{HZ} = 1\ AU\ (L_{star}/L_{Sun})^{0.5}$, and finally

$$a_{HZ} = 1\ AU \times ((L/L_{sun})/S_{eff})^{0.5}$$

This formula assumes that the planet has a similar albedo to Earth, that it rotates or redistributes the insolation as on Earth, and that it has a similar greenhouse effect. $S_{eff}$ is 1.90, 1.41, 1.05 and 1.05 for F, G, K and M stars respectively for the inner edge of the HZ (where runaway greenhouse occurs) and 0.46, 0.36, 0.27 and 0.27 for F, G, K and M stars respectively for the outer edge of the HZ (assuming a maximum greenhouse effect in the planet's atmosphere) (Kasting et al. 1993). On an Earth-like planet where the carbonate-silicate cycle is at work, the level of $CO_2$ in the atmosphere depends on the orbital distance: $CO_2$ is a trace gas close to the inner edge of the HZ but a major compound in the outer part of the HZ (Forget et al. 1997) (figure 5).

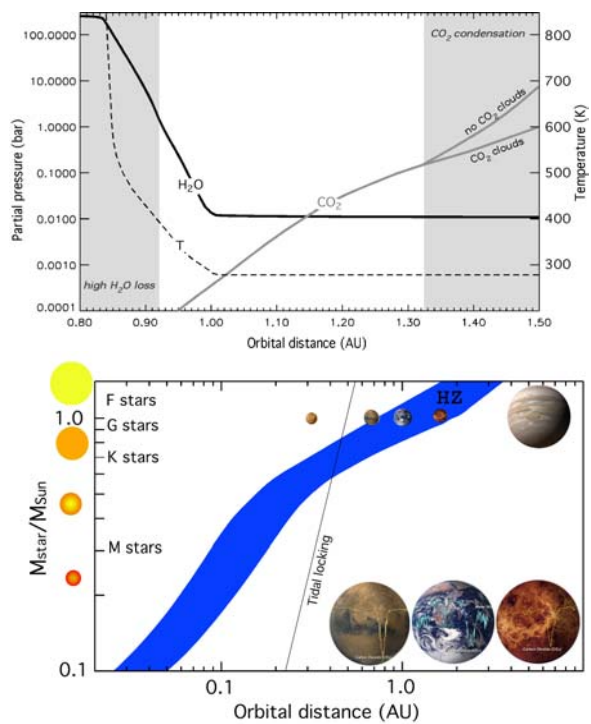

**Fig. 5: Surface conditions on a habitable planet within the habitable zone (data adapted from Kasting et al. 1993, Forget and Pierrhumbert 1997)(Kaltenegger & Selsis 2007) (top) the Habitable Zone as a function of stellar type (bottom panel).**

Earth-like planets close to the inner edge are expected to have a water-rich atmosphere or to have lost their water reservoir to space. This is one of the first theories we can test with a first generation space mission. However, the limits of the HZ are known qualitatively, more than quantitatively. This uncertainty is mainly due to the complex role of clouds and three-dimensional climatic effects not yet included in the modeling. Thus, planets slightly outside the computed HZ could still be habitable, while planets at habitable orbital distance may not be habitable because of their size or chemical composition.

As the HZ is defined for surface conditions only, chimio-lithotrophic life, which metabolism does not depend on the stellar light, can still exist outside the HZ, thriving in the interior of the planet where liquid water is available. Such metabolisms (at least the ones we know on Earth) do not produce $O_2$ and relies on very limited sources of energy (compared to stellar light) and electron donors (compared to $H_2O$ on Earth). They mainly catalyze reactions that would occur at a slower rate in purely abiotic conditions and they are thus not expected to modify a whole planetary environment in a detectable way.

### 2.1 Potential Biomarkers

Owen (Owen 1980) suggested searching for $O_2$ as a tracer of life. Oxygen in high abundance is a promising bio-indicator. Oxygenic photosynthesis, which by-product is molecular oxygen extracted from water, allows terrestrial plants and photosynthetic bacteria (cyanobacteria) to use abundant $H_2O$, instead of having to rely on scarce supplies of electron donor to reduce $CO_2$, like $H_2$ and $H_2S$. With oxygenic photosynthesis, the production of the biomass becomes limited only by nutriments and no longer by energy (light in this case) nor by the abundance of electron donors. Oxygenic photosynthesis at a planetary scale results in the storage of large amounts of radiative energy in chemical energy, in the form of organic matter. For this reason, oxygenic photosynthesis had a tremendous impact on biogeochemical cycles on Earth and eventually resulted in the global transformation of Earth environment. Less than 1ppm of atmospheric $O_2$ comes from abiotic processes (Walker 1977). Cyanobacteria and plants are responsible for this production by using the solar photons to extract hydrogen from water and using it to produce organic molecules from $CO_2$. This metabolism is called oxygenic photosynthesis. The reverse reaction, using $O_2$ to oxidize the organics produced by photosynthesis, can occur abiotically when organics are exposed to free oxygen, or biotical by eukaryotes breathing $O_2$ and consuming organics.

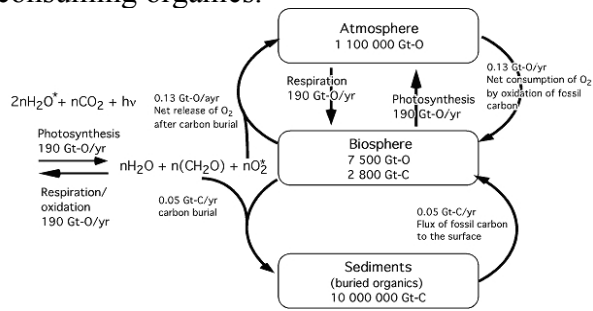

**Fig. 6: Oxygen Cycle on Earth**

Because of this balance, the net release of $O_2$ in the atmosphere is due to the burial of organics in sediments. Each reduced carbon buried results in a free $O_2$ molecule in the atmosphere. This net release rate is also balanced by weathering of fossilized carbon when exposed to the surface (see figure 6). The oxidation of reduced volcanic gasses such as $H_2$, $H_2S$ also accounts for a significant fraction of the oxygen losses. The atmospheric oxygen is recycled through respiration and photosynthesis in less than 10 000 yrs. In the case of a total extinction of Earth biosphere, the atmospheric $O_2$ would disappear in a few million years.

Reduced gases and oxygen have to be produced concurrently to be detectable in the atmosphere, as they react rapidly with each other. Thus, the chemical imbalance traced by the simultaneous signature of $O_2$ and/or $O_3$ and of a reduced gas like $CH_4$ can be considered as a signature of biological activity (Lovelock 1975). The spectrum of the Earth has exhibited a strong infrared signature of ozone for more than 2 billion years, and a strong visible signature of $O_2$ for an undetermined period of time between 2 and 0.8 billion years (depending on the required depth of the band for detection and also the actual evolution of the $O_2$ level) (Kaltenegger et al. 2007). This difference is due to the fact that a saturated ozone band appears already at very low levels of $O_2$ ($10^{-4}$ ppm) while the oxygen line remains unsaturated at values below 1 PAL (Segura et al. 2003). In addition, the stratospheric warming decreases with the abundance of ozone, making the $O_3$ band deeper for an ozone layer less dense than in the present atmosphere. The depth of the saturated $O_3$ band is determined by the temperature difference between the surface-clouds continuum and the ozone layer.

$N_2O$ is produced in abundance by life but only in negligible amounts by abiotic processes. Nearly all of Earth's $N_2O$ is produced by the activities of anaerobic denitrifying bacteria. $N_2O$ would be hard to detect in Earth's atmosphere with low resolution, as its abundance is low at the surface (0.3 ppmv) and falls off rapidly in the stratosphere. Spectral features of $N_2O$ would become more apparent in atmospheres with more $N_2O$ and/or less $H_2O$ vapor. Segura et al. (Segura et al. 2003) have calculated the level of $N_2O$ for different $O_2$ levels and found that, although $N_2O$ is a reduced species compared to $N_2$, its levels decreases with $O_2$. This is due to the fact a decrease in $O_2$ produces an increase of $H_2O$ photolysis resulting in the production of more hydroxyl radicals (OH) responsible for the destruction of $N_2O$.

The methane found in the present atmosphere of the Earth has a biological origin, but for a small fraction produced abiotically in hydrothermal systems where hydrogen is released by the oxidation of Fe by $H_2O$, and reacts with $CO_2$. Depending on the degree of oxidation of a planet's crust and upper mantle, such non-biological mechanisms can also produce large amounts of $CH_4$ under certain circumstances. Therefore, the detection of methane alone cannot be considered a sign of life, while its detection in an oxygen-rich atmosphere would be difficult to explain in the absence of a biosphere. Note that methane on Mars may have been detected (Mumma et al. 2009) while the atmosphere of Mars contains 0.1% of $O_2$ and some ozone. In this case, the amounts involved are extremely low and the origin of the Martian $O_2$ and $O_3$ is known to be photochemical reactions initiated by the photolysis of $CO_2$ and water vapor. If confirmed, the presence of methane could be explained by subsurface geochemical process, assuming that reducing conditions exist on Mars below the highly oxidized surface. The case of $NH_3$ is similar to the one of $CH_4$. They are both released into Earth's atmosphere by the biosphere with similar rates but the atmospheric level of $NH_3$ is orders of magnitude lower due to its very short lifetime under UV irradiation. The detection of $NH_3$ in the atmosphere of a habitable planet would thus be extremely interesting, especially if found with oxidized species. The detection of $H_2O$ and $CO_2$, not as biosignatures themselves, are important in the search for signs of life because they are raw materials for life and thus necessary for planetary habitability.

There are other molecules that could, under some circumstances, act as excellent biomarkers, e.g., the manufactured chloro-fluorocarbons ($CCl_2F_2$ and $CCl_3F$) in our current atmosphere in the thermal infrared waveband, but their abundances are too low to be spectroscopically observed at low resolution.

### 2.1.1 Low resolution spectral information in the visible to near-IR

In the visible to near-infrared one can see increasingly strong $H_2O$ bands at 0.73 μm, 0.82 μm, 0.95 μm, and 1.14 μm. The strongest $O_2$ feature is the saturated Frauenhofer A-band at 0.76 μm. A weaker feature at 0.69 μm can not be seen with low resolution (see figure 3). $O_3$ has a broad feature, the Chappuis band, which appears as a broad triangular dip in the middle of the visible spectrum from about 0.45 μm to 0.74 μm. The feature is very broad and shallow. Methane at present terrestrial abundance (1.65 ppm) has no significant visible absorption features but at high abundance, it has strong visible bands at 0.88 μm, and 1.04 μm, readily detectable e.g. in early Earth models (see figure 7). $CO_2$ has negligible visible features at present abundance, but in a high $CO_2$-atmosphere of 10% $CO_2$, like in an early Earth evolution stage, the weak 1.06 μm band could be observed. In the UV $O_3$ shows a strong feature, not discussed here. The red edge of land plants developed about 0.44Ga. It could be observed on a cloud-less Earth or if the cloud pattern is known (see section 3).

### 2.1.2 Low resolution spectral information in the mid-IR

In the mid-IR on Earth the detectable signatures of biological activity in low resolution are the combined detection of 9.6 μm $O_3$ band, the 15 μm $CO_2$ band and the 6.3 μm $H_2O$ band or its rotational band that extends from 12 μm out into the microwave region (Selsis 2003). The 9.6 μm $O_3$ band is highly saturated and is thus a poor quantitative indicator, but an excellent qualitative indicator for the existence of even traces of $O_2$. $CH_4$ is not readily identified using low resolution spectroscopy for present-day Earth, but the methane feature at 7.66 μm in the IR is easily detectable at higher abundances (see e.g. 100x on early Earth (Kaltenegger et al. 2007)), provided of course that the spectrum contains the whole band and a high enough SNR. Taken together with molecular oxygen, abundant $CH_4$ can indicate biological processes (see also (Sagan et al. 1993)(Segura et al. 2003)). Although methane's abundance is less than 1 ppm in Earth atmosphere, the 7.75 μm shows up in a medium resolution (Res=100) infrared spectrum. Three $N_2O$ features in the thermal infrared are detectable at 7.75 μm and 8.52 μm, and 16.89 μm for levels higher than in the present atmosphere of the Earth.

## 3 Geological evolution, Cryptic worlds, Abiotic sources, and Host stars

### 3.1 Evolution of biomarkers over geological times on Earth

One crucial factor in interpreting planetary spectra is the point in the evolution of the atmosphere when its biomarkers and its habitability become detectable.

The spectrum of the Earth has not been static throughout the past 4.5 Ga. This is due to the variations in the molecular abundances, the temperature structure, and the surface morphology over time. At about 2.3 Ga oxygen and ozone became abundant, affecting the atmospheric absorption component of the spectrum. At about 0.44 Ga, an extensive land plant cover followed, generating the red chlorophyll edge in the reflection spectrum. The composition of the surface (especially in the visible), the atmospheric composition, and temperature-pressure profile can all have a significant influence on the detectabilty of a signal. Figure 7 shows theoretical visible and mid-infrared spectra of the Earth at six epochs during its geological evolution (Kaltenegger et al. 2007). The epochs are chosen to represent major developmental stages of the Earth, and life on Earth. If an extrasolar planet is found with a corresponding spectrum, we can use the stages of evolution of our planet to characterizing it, in terms of habitability and the degree to which it shows signs of life. Furthermore we can learn about the evolution of our own planet's atmosphere and possible the emergence of life by observing exoplanets in different stages of their evolution. Earth's atmosphere has experienced dramatic evolution over 4.5 billion years, and other planets may exhibit similar or greater evolution, and at different rates. It shows epochs that reflect significant changes in the chemical composition of the atmosphere. The oxygen and ozone absorption features could have been used to indicate the presence of biological activity on Earth anytime during the past 50% of the age of the solar system. Different signatures in the atmosphere are clearly visible over Earth's evolution and observable with low resolution.

Spectra of the Earth exploring temperature sensitivity (hot house and cold scenario) and different singled out stages of its evolution (e.g. (Pavlov et al. 2000)(Schindler & Kasting 2000)(Traub & Jucks 2002)) as well as explore the evolution of the expected spectra of Earth (Kaltenegger et al. 2007) produce a variety of spectral fingerprints for our own planet (see also Grenfell et al this volume).

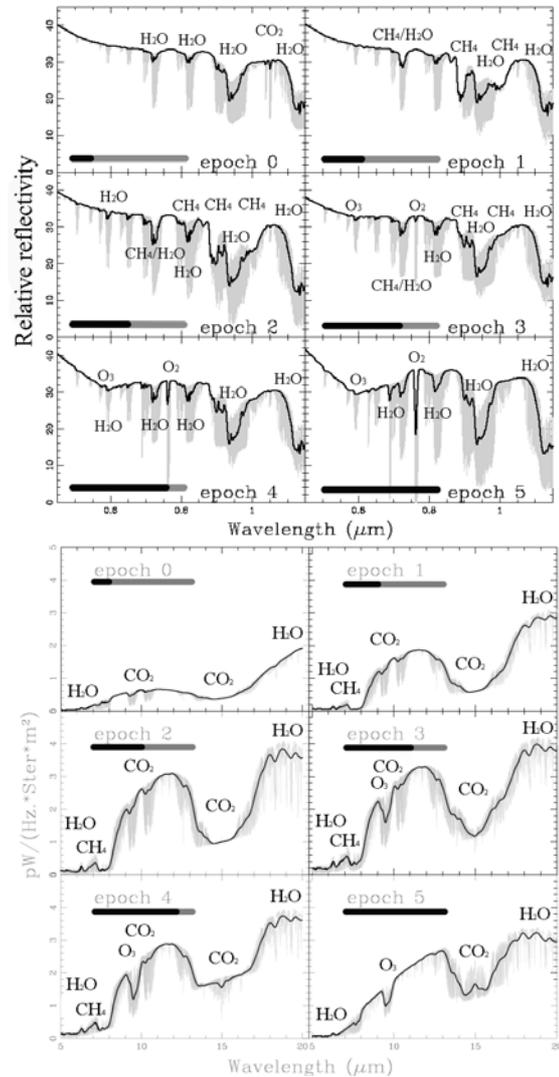

Fig. 7: The visible to near-IR (top) and mid IR (bottom) spectral features on an Earth-like planet change considerably over its evolution from a $CO_2$ rich (epoch 0) to a $CO_2/CH_4$-rich atmosphere (epoch 3) to a present-day atmosphere (epoch 5). The bold lines show spectral resolution of 80 and 25 comparable to the proposed visible TPF and Darwin/TPF-I mission concept, respectively.

Those spectra will be used as part of a big grid to characterize any exoplanets found and influences the design requirements for a spectrometer to detect habitable planets (Kaltenegger et al. 2007).

### 3.2 Abiotic sources of biomarkers

Abiotic sources of biomarkers are very important to assess, so that we can identify when it might constitute a "false positive" for life. $CH_4$ is an abundant constituent of the cold planetary atmospheres in the outer solar system. On Earth, it is produced abiotically in hydrothermal systems where $H_2$ (produced from the oxidation of Fe by water) reacts with $CO_2$ in a certain range of pressures and temperatures. In the absence of atmospheric oxygen, abiotic methane could build up to detectable levels. Therefore, the detection of $CH_4$ cannot be attributed unambiguously to life.

$O_2$ also has abiotic sources, the first one is the photolysis of $CO_2$, followed by recombination of O atoms to form $O_2$ (O + O + M → $O_2$ + M), a second one is the photolysis of $H_2O$ followed by escape of hydrogen to space. The first source is a steady state maintained by the stellar UV radiation, but with a constant elemental composition of the atmosphere while the second one is a net source of oxygen. In order to reach detectable levels of $O_2$ (in the reflected spectrum), the photolysis of $CO_2$ has to occur in the absence of outgassing of reduced species and in the absence of liquid water, because of the wet deposition of oxidized species. Normally, the detection of the water vapor bands simultaneously with the $O_2$ band can rule out this abiotic mechanism (Segura et al. 2007), although one should be careful, as the vapor pressure of $H_2O$ over a high-albedo icy surface might be high enough to produce detectable $H_2O$ bands. In the infrared, this process cannot produce a detectable $O_3$ feature (Selsis et al. 2002). The loss of hydrogen to space can result in massive oxygen leftovers: more than 200 bars of oxygen could build up after the loss of the hydrogen contained in the Earth Ocean. However, the case of Venus tells us that such oxygen leftover has a limited lifetime in the atmosphere (because of the oxidation of the crust and the loss of oxygen to space): we do not find $O_2$ in the Venusian atmosphere despite the massive loss of water probably experienced in the early history of the planet. Also, such evaporation-induced build up of $O_2$ should occur only closer to a certain distance from the Star and affect small planets with low gravity more dramatically. For small planets (<0.5 $M_{Earth}$) close to inner edge of the habitable zone (<0.93 AU from the present Sun), there is a risk of abiotic oxygen detection, but this risk becomes negligible for big planets further away from their star. The fact that, on the Earth, oxygen and indirectly ozone are by-products of the biological activity does not mean that life is the only process able to enrich an atmosphere with these compounds. The question of the abiotic synthesis of biomarkers is crucial, but only very few studies have been dedicated to it (Rosenqvist & Chassefiere 1995)(Leger et al. 1993)(Lagrange et al. 2009)(Selsis et al. 2002).

### 3.3 Cryptic worlds, Surface- and Cloud-features

While they efficiently absorb the visible light, photosynthetic plants have developed strong infrared reflection (possibly as a defense against overheating and chlorophyll degradation) resulting in a steep change in reflectivity around 700 nm, called the red-edge. The primary molecules that absorb the energy and convert it to drive photosynthesis ($H_2O$ and $CO_2$ into sugars and $O_2$) are chlorophyll A (0.450 μm) and B (0.680 μm). The exact wavelength and strength of the spectroscopic "vegetation red edge" (VRE) depends on the plant species and environment. On Earth around 440 million years ago (Pavlov et al. 2003)(Schopf 1993), an extensive land plant cover developed, generating the red chlorophyll edge in the reflection spectrum between 700 and 750nm. Averaged over a spatially unresolved hemisphere of Earth, the additional reflectivity of this spectral feature is typically only a few percent (see also (Montanes-Rodriguez et al. 2005)(Kaltenegger & Traub 2009)). Several groups (Arnold et al. 2002)(Christensen et al. 1997)(Montanes-Rodriguez et al. 2007)(Woolf et al. 2002)(Turnbull et al.

2006) have measured the integrated Earth spectrum via the technique of Earthshine, using sunlight reflected from the non-illuminated, or "dark", side of the moon. Earthshine measurements have shown that detection of Earth's VRE is feasible if the resolution is high and the cloud coverage is known, but is made difficult owing to its broad, essentially featureless spectrum and cloud coverage.

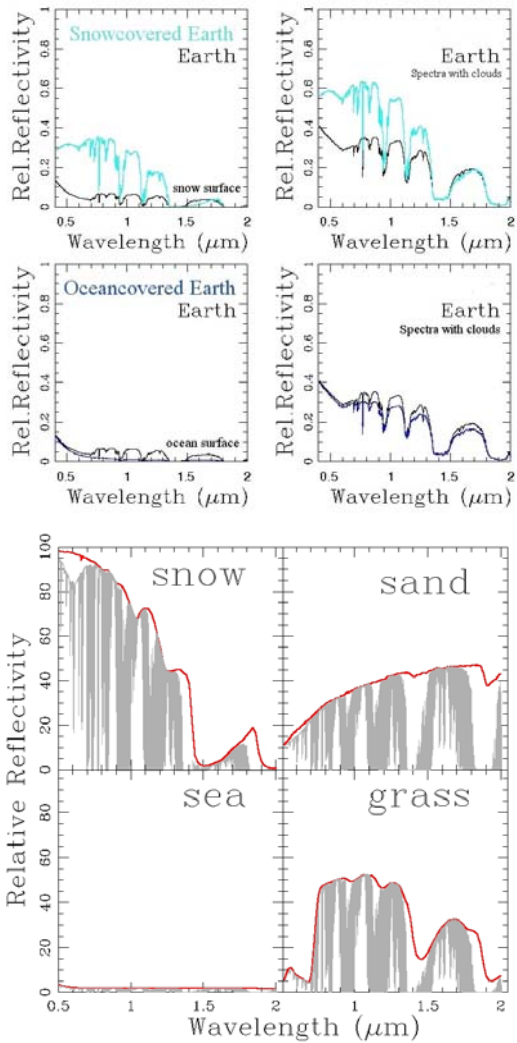

**Fig. 8: (top) Spectra of present-day Earth with a total ocean and snow cover without (left) and with (right) clouds for a disk averaged view. Note that the low albedo of the ocean reduces the overall flux while the high albedo of snow reflects more sunlight of the planets surface. (bottom panel) Reflectivity of different surfaces for present-day cloud-free Earth atmosphere.**

Our knowledge of the reflectivity of different surface components on Earth - like deserts, ocean and ice - helps in assigning the VRE of the Earthshine spectrum to terrestrial vegetation. Earth's hemispherical integrated vegetation red-edge signature is very weak, but planets with different rotation rates, obliquities, land-ocean fraction, and continental arrangement may have lower cloud-cover and higher vegetated fraction (see e.g. (Seager & Ford 2002)). Knowing that other pigments exist on Earth and that some minerals can exhibit a similar spectral shape around 750 nm (Seager et al. 2005), the detection of the red-edge of the chlorophyll on exoplanets, despite its interest, will not be unambiguous.

Picking the most different reflecting surfaces (snow with a high albedo and sea with an extremely low albedo), figure 8 shows the maximum effect surface coverage could have on the amount of light reflected from an exoplanet – assuming the whole planet surface is covered with that one material, the surface area is the same, and also artificially assuming similar cloud coverage and atmospheres for comparison. Assuming that similar photosynthesis would evolve on a planet around other stellar types, possible different types of spectral signature have been modeled (Tinetti et al. 2006) that could be a guide to interpreting other spectral signature. Those signatures will be difficult to verify through remote observations as being of biological origin.

On the Earth, photosynthetic organisms are responsible for the production of nearly all of the oxygen in the atmosphere. However, in many regions of the Earth, and particularly where surface conditions are extreme, for example in hot and cold deserts, photosynthetic organisms can be driven into and under substrates where light is still sufficient for photosynthesis. These communities exhibit no detectable surface spectral signature. The same is true of the assemblages of photosynthetic organisms at more than a few meters depth in water bodies. These communities are widespread and dominate local photosynthetic productivity. Figure 9 shows known cryptic photosynthetic communities and their calculated disk-averaged spectra of such hypothetical cryptic photosynthesis worlds.

Such world is Earth-analogs that would not exhibit a biological surface feature in the disc-averaged spectrum (Cockell et al. 2009).

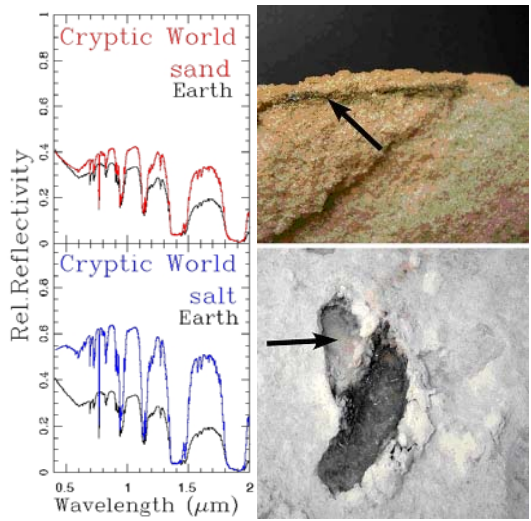

Fig. 9: Two examples of spectra of land-based cryptic photosynthetic communities. (top) a cryptoendolithic lichen (arrow) inhabiting the interstices of sandstone in the Dry Valleys of the Antarctic, (bottom) endoevaporites inhabit a salt crust visible as pink pigmentation (arrow) (photo: Marli Bryant Miller) and their respective calculated clear reflection spectra. Substrates represent typical habitats for different cryptic biota (Cockell et al 2009).

Another topic that has been proposed to discover continents and seas on an exoplanet is the daily variation of the surface albedo in the visible (Ford et al. 2001)(Seager & Ford 2002)(Palle et al. 2007). On a cloud-free Earth, the diurnal flux variation in the visible caused by different surface features rotating in and out of view could be high, assuming hemispheric inhomogeneity. When the planet is only partially illuminated, a more concentrated signal from surface features could be detected as they rotate in and out of view on a cloudless planet (William et al. 2008). Earth has an average of 60% cloud coverage, what prevents easy identification of the features without knowing the cloud distribution. Clouds are an important component of exoplanet spectra because their reflection is high and relatively flat with wavelength.

Clouds reduce the relative depths, full widths, and equivalent widths of spectral features, weakening the spectral lines in both the thermal infrared and visible (Kaltenegger et al. 2007). In the thermal infrared, clouds emit at temperatures that are generally colder than the surface, while in the visible the clouds themselves have different spectrally-dependent albedo that further influence the overall shape of the spectrum.

If one could record the planet's signal with a very high time resolution (a fraction of the rotation period of the planet) and SNR, one could determine the overall contribution of clouds to the signal (Cowan et al 2009)(Palle et al 2007). During each of these individual measurements, one has to collect enough photons for a high individual SNR per measurement to be able to correlate the measurements to the surface features, what precludes this method for first generation missions that will observe a minimum of several hours to achieve a SNR of 5 to 10. For Earth (Cowan et al 2009)(Palle et al 2007) these measurements show a correlation to Earth's surface feature because the individual measurements are time resolved as well as have an individual high SNR, making it a very interesting concept for future generations of missions.

### 3.4 Influence of host-stars

The range of characteristics of planets is likely to exceed our experience with the planets and satellites in our own Solar System by far. Models of planets more massive than our Earth – rocky SuperEarths - need to consider the changing atmosphere structure, as well as the interior structure of the planet (see e.g.(Seager et al. 2007)(Valencia et al. 2007)). Also, Earth-like planets orbiting stars of different spectral type might evolve differently (Selsis 2000)(Segura et al. 2003)(Segura et al. 2005). Modeling these influences will help to optimize the design of the proposed instruments to search for Earth-like planets. The spectral resolution required for optimal detection of habitability and biosignatures has to be able to detect those features on our own planet for the dataset we have over its evolution.

Using a numerical code that simulates the photochemistry of a wide range of planetary atmospheres several groups (Selsis 2000, Segura et al. 2003, 2005, Paillet 2006, Grenfell et al. 2007) have simulated a replica of our planet orbiting different types of star: F-type star (more massive and hotter than the Sun) and a K-type star (smaller and cooler than the Sun). The models assume same background composition of the atmosphere as well as the strength of biogenic sources.

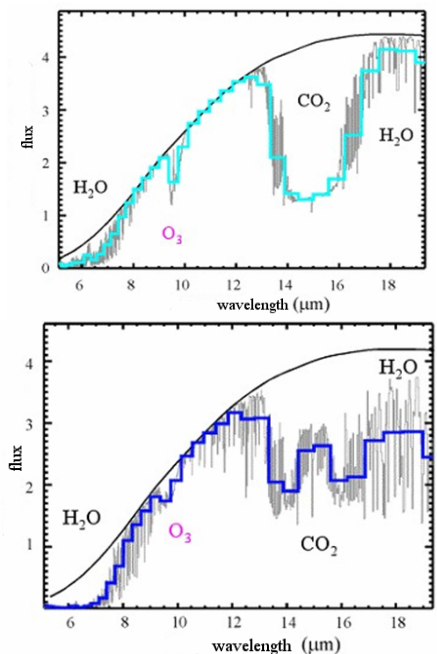

**Fig. 10: Calculated IR spectrum of an Earth-analog with resolution of 30 around an F star (top) and K star (bottom panel) (Selsis 2002).**

A planet orbiting a K star has a thin $O_3$ layer, compared to Earth's one, but still exhibits a deep $O_3$ absorption: indeed, the low UV flux is absorbed at lower altitudes than on Earth which results in a less efficient warming (because of the higher heat capacity of the dense atmospheric layers). Therefore, the ozone layer is much colder than the surface and this temperature contrast produces a strong feature in the thermal emission. The process works the other way around in the case of an F-type host star. Here, the ozone layer is denser and warmer than the terrestrial one, exhibiting temperatures about as high as the surface temperature. Thus, the resulting low temperature contrast produces only a weak and barely detectable feature in the infrared spectrum. This comparison shows that planets orbiting G (solar) and K-type stars may be better candidates for the search for the $O_3$ signature than planets orbiting F-type stars (see figure 10). This result is promising since G and K-type stars are much more numerous than F-type stars, the latter being rare and affected by a short lifetime (less than 1 Gyr).

## 4 Summary

Any information we collect on habitability, is only important in a context that allows us to interpret, what we find. To search for signs of life we need to understand how the observed atmosphere physically and chemically works. Knowledge of the temperature and planetary radius is crucial for the general understanding of the physical and chemical processes occurring on the planet. These parameters as well as an indication of habitability can be determined with low resolution spectroscopy and low photon flux, as assumed for first generation space missions. The combination of spectral information in the visible (starlight reflected off the planet) as well as in the mid-IR (planet's thermal emission) allows a confirmation of detection of atmospheric species, a more detailed characterization of individual planets but also to explore a wide domain of planet diversity. Being able to measure the outgoing shortwave and longwave radiation as well as their variations along the orbit, to determine the albedo and identify greenhouse gases, would in combination allow us to explore the climate system at work on the observed worlds, as well as probe planets similar to our own for habitable conditions.

The emerging field of extrasolar planet search has shown an extraordinary ability to combine research by astrophysics, chemistry, biology and geophysics into a new and exciting interdisciplinary approach to understand our place in the universe.

**Acknowledgement**
L. Kaltenegger acknowledges the support of the Harvard Origins of Life Initiative and the NASA Astrobiology Institute.